\documentclass[a4paper]{jpconf}
\newcommand{\GeVc}{\ensuremath{\textrm{GeV}/c}~}
\newcommand{\dEdx}{\ensuremath{\textrm{d}E/\textrm{d}x}}
\newcommand{\nsigmaTPC}{\ensuremath{n_{\sigma}^{\textrm{\tiny TPC}}}~}
\newcommand{\RpPb}{\ensuremath{R_{\rm pPb}}~}
\newcommand{\RPbPb}{\ensuremath{R_{\rm PbPb}}~}

\usepackage{graphicx}
\begin{document}
\title{Separation of the Charm and Beauty Production in p--Pb and Pb--Pb Collisions with ALICE}

\author{Martin V\"olkl for the ALICE Collaboration}

\address{Physikalisches Institut der Universit\"at Heidelberg, Heidelberg, Germany}

\ead{voelkl@physi.uni-heidelberg.de}

\begin{abstract}
Measurements of heavy (charm and beauty) quarks provide useful insights into the properties of the quark--gluon plasma. The good particle identification capabilities of ALICE make a measurement based on the electrons from semi-leptonic decays of heavy-flavour hadrons feasible. This approach makes use of the large branching ratios ($\approx 10-20\%$) of heavy--flavour hadrons into electrons. Separation of the contribution from beauty-hadron decay electrons was achieved by utilizing the large decay length of the associated hadrons, making the investigation of beauty quarks in the medium possible. By comparing measurements in p--Pb and Pb--Pb collisions, it is possible to disentangle effects of cold nuclear matter from those of the hot and dense medium. The results show a strong change in the transverse momentum distribution of electrons from beauty-hadron decays in central Pb--Pb collisions with respect to the proton--proton case. No significant change from proton--proton collisions can be seen in the p--Pb case, suggesting that the modification is a final state effect.
\end{abstract}

\section{Introduction}
The interaction of heavy quarks with the medium produced in an ultrarelativistic heavy-ion collision can yield insight into the properties of the quark--gluon plasma. The use of heavy quarks as probes has several advantages \cite{Prino:2016cni}, among them the defined moment of creation in the initial hard scatterings of the collision and the clear association of the heavy quark in the medium and the hadron it forms. In the interaction with the medium, heavy quarks with large momenta may loose energy to the medium, while those with low momenta can participate in the collective motion of the medium. Measurements of these processes give an insight into the properties of the medium. To assess the strength of the interaction, it is useful to compare the transverse momentum distribution of the resulting hadrons in central Pb--Pb collisions with that in pp collisions. In addition, it is also useful to compare to a measurement in p--Pb collisions to gain an insight into initial state effects due to the use of nuclei as projectiles, such as the modification of the parton distribution function.
A useful property of heavy-flavour hadrons is the typically large branching ratio of decays with an electron in the final state. As a result most electrons at mid-rapidity and with a momentum of a few \GeVc come from heavy-flavour sources. With the excellent particle identification capabilities of the ALICE detector, an approach based on the measurement of electrons becomes feasible. The comparatively large decay length of beauty hadrons ($c \tau \approx 500 \, \mu\rm{m}$) can be utilized to separate the contribution from beauty from that of charm and other background sources. The measurements can broadly be split into two main steps: Identifying a clean sample of electrons from all measured particles and secondly identifying the contribution from beauty-hadron decays.

\section{Electron Identification}

The Pb--Pb data set discussed here was recorded by the ALICE detector setup in 2010 \cite{Aamodt:2008zz,Abelev:2014ffa}, while the p--Pb data set was recorded in 2013. Electrons were identified using the signal of the Time Projection Chamber (TPC) and Time-of-Flight detector (TOF). The time of flight was required to be compatible with the expectation for an electron within $3\sigma$. The TPC measures the energy deposit of particles in the gas (\dEdx). The signal of the main hadronic background -- charged pions -- lies below the electron signal, suggesting an asymmetric selection range which was set to be $-0.5<\nsigmaTPC <3$, where \nsigmaTPC is the deviation of the signal to an electron hypothesis in units of the detector resolution.

\begin{figure}[th]
\begin{minipage}{17pc}
\includegraphics[width=1.\textwidth]{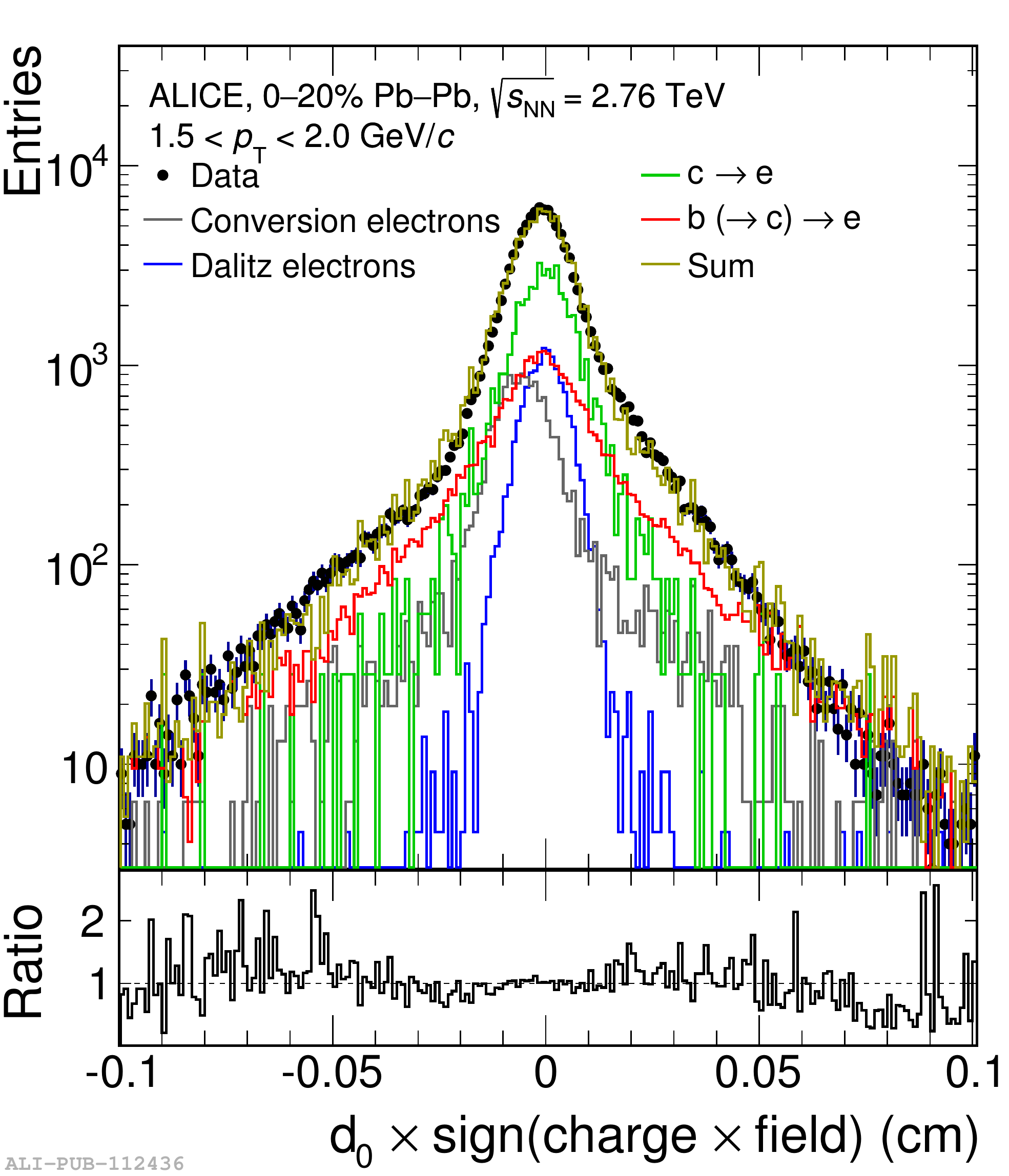}
\caption{\label{IPDist}Impact parameter distributions in Pb--Pb collisions \cite{OurBeautyPaper}.}
\end{minipage} 
\hspace{1pc}
\begin{minipage}{19pc}
\includegraphics[width=1.\textwidth]{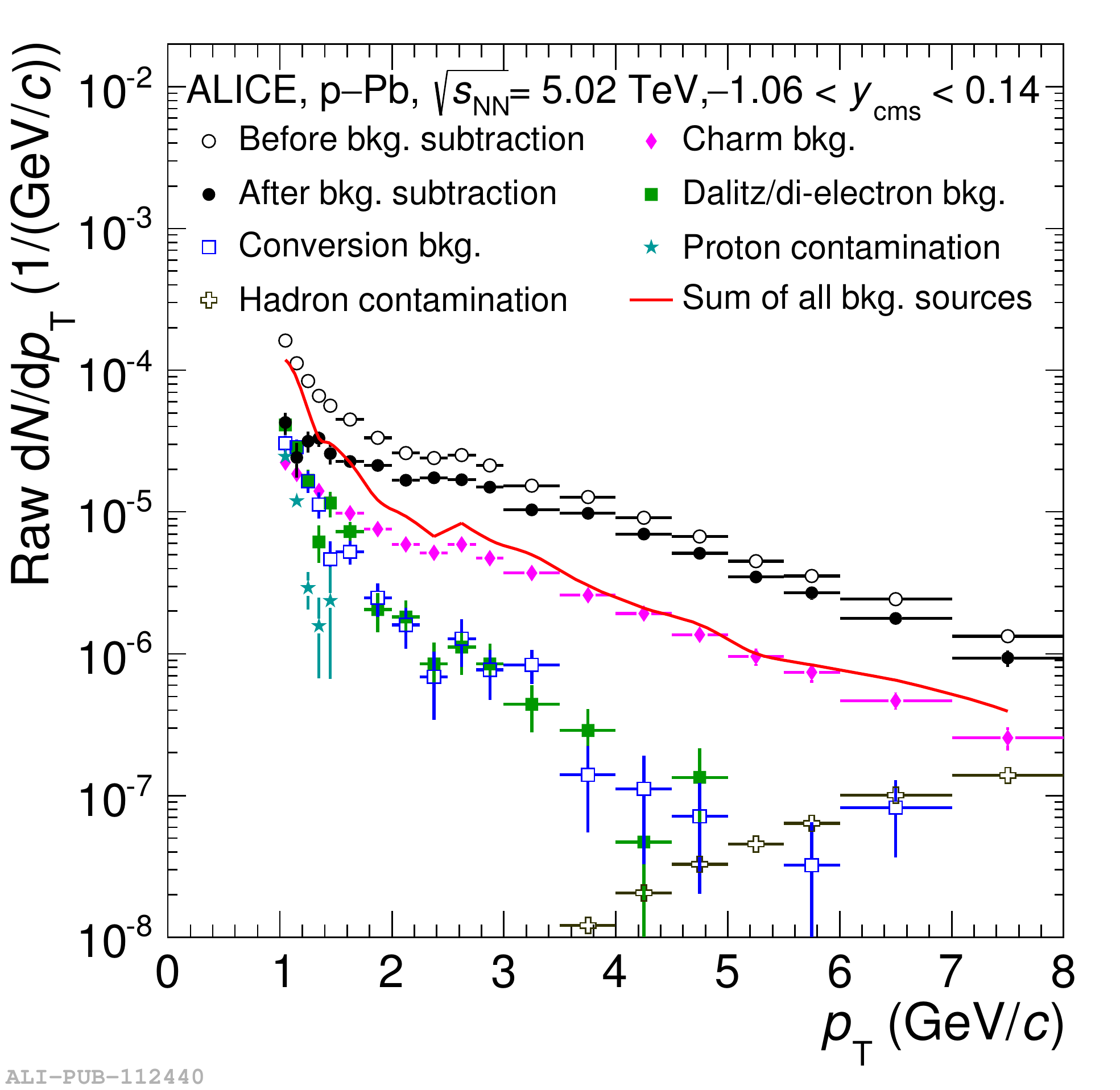}
\caption{\label{pPbContributions}Background contributions in the p--Pb electron sample with applied impact parameter requirement \cite{OurBeautyPaper}.}
\end{minipage}\hspace{2pc}%

\end{figure}

\section{Signal Extraction}

The signal electrons can come from the decay of the beauty hadrons directly or from the decay of an intermediate charm state. A differentiating property of the beauty--hadron decays is the fairly large decay length of the beauty hadrons. From the single electron tracks it is not possible to deduce the decay length directly. An indirect way is the use of the track impact parameter. The impact parameter is obtained by tracing the measured track back towards the reconstructed interaction vertex. If the electron is produced at a secondary vertex, then this will typically not lead to the primary vertex. The impact parameter is the minimum distance to the primary vertex in the transverse plane. In the definition used for the analysis, it has a positive or negative sign depending on whether the interaction vertex is inside or outside of the circle the track describes in the transverse plane. Electrons from beauty-hadron decays will typically have a larger (absolute) impact parameter value than electrons from the background sources.

To discuss the different electron sources, it is useful to group them by the shape of their impact parameter distribution. The sources can be separated into the following four categories, which are also shown in Fig. \ref{IPDist}:
\begin{description}
\item[Electrons from beauty-hadron decays.] Due to the large decay length ($c \tau \approx 500 ~\mathrm{\mu m}$) of beauty hadrons this distribution is the widest among the different electron sources.
\item[Electrons from charm-hadron decays.] The smaller decay length of the charm hadrons ($c \tau \approx 100-300~ \mathrm{\mu m}$) leads to a narrower distribution as compared to electrons from beauty-hadron decays.
\item[Electrons from photon conversions in the detector material.]  Due to the small opening angle of photon conversions, a non-zero impact parameter appears only due to the influence of the magnetic field. With the definition given above, this leads to an asymmetric impact parameter distribution with a maximum at negative values.
\item[All remaining electron sources.] This contribution is dominated by electrons from the decay of light hadrons with a small contribution from the decay of particles with strangeness. Many of these electrons come from Dalitz decays of $\pi^{0}$ and they will thus be referred to as Dalitz electrons in the following.
\end{description}

The approach of the analysis for p--Pb collisions is to require a minimum (absolute) impact parameter for the measured electrons. This strongly reduces the background. The remaining background can then be subtracted. After a correction for the associated efficiency, this leads to an estimate for the number of electrons from beauty-hadron decays. The estimate for the background contributions comes from the information of other measurements by ALICE, which are corrected with the corresponding efficiency. The resulting contributions are shown in Fig. \ref{pPbContributions}.

For the Pb--Pb analysis, the information about the different contributions is obtained by fitting the full impact parameter distribution as shown in Fig. \ref{IPDist}. For this purpose, estimates for the distributions are added with a free amplitude parameter to reproduce the full distribution. The information about these distributions comes from Monte Carlo simulations of the event properties and the detector setup. These have finite statistics, which has to be taken into account in the fitting procedure \cite{Barlow1993219}.

In both cases, the accurate reproduction of the impact parameter distributions by the simulations has to be carefully assessed and corrected where necessary \cite{OurBeautyPaper}.

\begin{figure}[th]
\begin{minipage}{16pc}
\includegraphics[width=1.\textwidth]{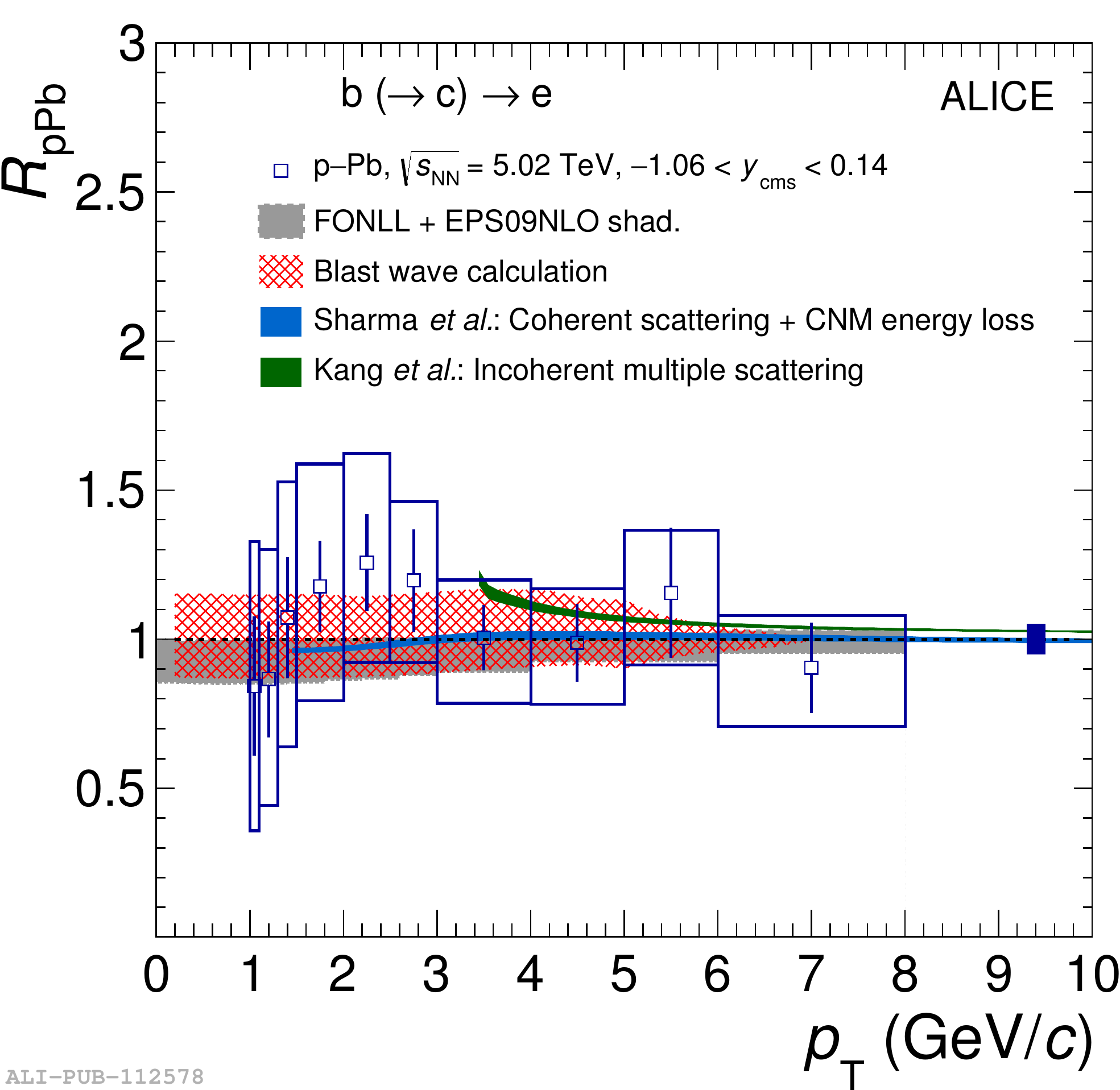}
\caption{\label{RpPb} Measured \RpPb compared with models \cite{OurBeautyPaper}.}
\end{minipage}\hspace{2pc}%
\begin{minipage}{16pc}
\includegraphics[width=1.\textwidth]{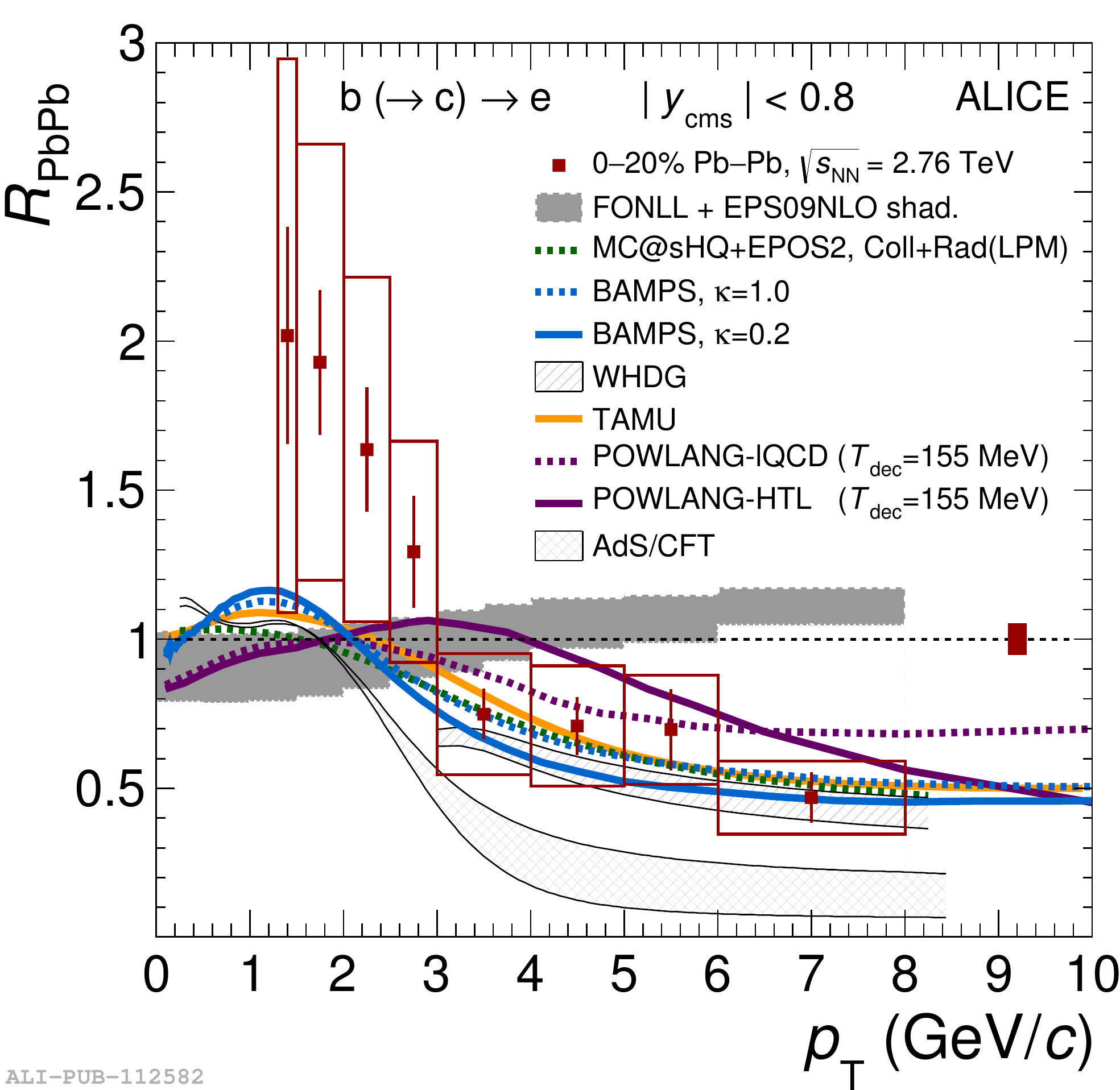}
\caption{\label{RAA} Measured \RPbPb in comparison with models \cite{OurBeautyPaper}.}
\end{minipage} 
\end{figure}

\section{Results}

The comparison of the results was done using the nuclear modification factor, which compares the $p_{\rm T}$-distribution of the electrons per nuclear collision to the $p_{\rm T}$--distribution of electrons in an equivalent number of proton--proton collisions \cite{Abelev:2012sca}:

\[R_\mathrm{AA}(p_{\mathrm{T}})=\frac{1}{\langle N_{\mathrm{coll}}\rangle} \frac{\mathrm{d} N_\mathrm{AA}/\mathrm{d} p_\mathrm{T}}{\mathrm{d} N_\mathrm{pp}/\mathrm{d} p_\mathrm{T}}=\frac{1}{\langle T_{\mathrm{AA}}\rangle} \frac{\mathrm{d} N_\mathrm{AA}/\mathrm{d} p_\mathrm{T}}{\mathrm{d} \sigma_\mathrm{pp}/\mathrm{d} p_\mathrm{T}} ~ ,\]
where $N_{\mathrm{coll}}$ is the expected number of nucleon-nucleon collisions and $T_{\mathrm{AA}}$ is the nuclear overlap function. The resulting ratio should be unity if a heavy-ion collision is equivalent to a superposition of independent nucleon-nucleon collisions.

In p--Pb collisions (Fig. \ref{RpPb}) the nuclear modification factor is consistent with unity and also in agreement with the expectations from theoretical models. This means, that no large deviation from unity due to such effects is expected in the \RPbPb.

The \RPbPb shown in Fig. \ref{RAA} is below unity above $p_{\rm{T}} = 3 \, \rm{GeV}/c$ and rises towards lower momenta. Given that cold nuclear matter effects are expected to be small, this suggests that fast quarks in the medium are slowed down by the interaction with the medium. Most of the models based on in-medium energy loss show a similar trend and also reasonable agreement with each other \cite{OurBeautyPaper}. The AdS/CFT-inspired model \cite{Horowitz:2015dta} underestimates the nuclear modification factor, while the POWLANG model \cite{Alberico:2011zy} based on hard thermal loops gives a higher estimate. In conclusion, modification of the $p_{\rm{T}}$-distribution of electrons from heavy-flavour hadron decays at mid-rapidity is caused by the interaction with the hot and dense medium. The magnitude of the effects can be reproduced by several model calculations.

\section*{References}

\bibliography{ThesisBib.bib}

\end{document}